\documentclass[a4paper,twocolumn,pre,aps,10pt]{revtex4-1}
\usepackage[pdftex]{graphicx}
\usepackage{bm}
\usepackage{amsfonts}
\usepackage{amsmath}
\usepackage{amssymb}
\usepackage{wasysym}
\usepackage{marvosym}
\usepackage{amsbsy}
\usepackage{color}
\usepackage{array}
\usepackage{dcolumn}
\usepackage{upgreek}
\usepackage{sidecap}
\usepackage{natbib}
\usepackage{float}
\usepackage{blindtext}

\definecolor{amber}{rgb}{1.0, 0.75, 0.0}
\definecolor{arsenic}{rgb}{0.23, 0.27, 0.29}
\definecolor{battleshipgrey}{rgb}{0.52, 0.52, 0.51}
\definecolor{charcoal}{rgb}{0.21, 0.27, 0.31}
\definecolor{darkelectricblue}{rgb}{0.33, 0.41, 0.47}
\definecolor{firebrick}{rgb}{0.7, 0.13, 0.13}

\begin{document}

\title{Bacterial Footprints in Elastic Pillared Microstructures}


\author{Arturo Susarrey-Arce}
\email{a.susarreyarce@utwente.nl}
\author{Jos\'e Federico Hern\'andez-S\'anchez}
\author{Marco Marcello}
\author{Yuri Diaz-Fernandez}
\author{Alina Oknianska}
\author{Ioritz Sorzabal-Bellido}
\author{Roald Tiggelaar}
\author{Detlef Lohse}
\author{Han Gardeniers}
\author{Jacco Snoeijer}
\author{Alvaro Marin}
\email{a.marin@utwente.nl}
\author{Rasmita Raval}

\affiliation{$^1$ Open Innovation Hub for Antimicrobial Surfaces at the Surface Science Research Centre and Department of Chemistry, University of Liverpool, United Kingdom.\\
$^2$ King Abdullah University of Science and Technology, Division of Physical Sciences and Engineering and Clean Combustion Research Center, Saudi Arabia.\\
$^3$ Institute of Integrative Biology, University of Liverpool, Biosciences Building, United Kingdom.\\
$^4$School of Health Sciences, Liverpool Hope University, Hope Park, United Kingdom.\\
$^5$ NanoLab Cleanroom, MESA+ Institute for Nanotechnology, University of Twente, The Netherlands.\\
$^6$ Physics of Fluids Group, MESA+ Institute for Nanotechnology, University of Twente, The Netherlands.\\
$^7$ Mesoscale Chemical Systems, MESA+ Institute for Nanotechnology, University of Twente, The Netherlands.}

\begin{abstract}
Soft substrates decorated with micropillar arrays are known to be sensitive to deflection due to capillary action. In this work, we demonstrate micropillared epoxy surfaces are sensitive to single drops of bacterial suspensions. The micropillars can show significant deformations upon evaporation, just as capillary action does in soft substrates. The phenomenon has been studied with five bacterial strains {\textit{S. epidermidis, L. sakei, P. aeruginosa, E. coli}} and {\textit{B. subtilis}}. The results reveal that only droplets containing motile microbes with flagella stimulate micropillar bending, which leads to significant distortions and pillar aggregations forming dimers, trimers, and higher order clusters. Such deformation is manifested in characteristic patterns that are left on the microarrayed surface following evaporation, and can be easily identified even by the naked eye. Our findings could lay the ground for the design and fabrication of mechanically responsive substrates, sensitive to specific types of  microorganisms.
\end{abstract}

\maketitle

\section{Introduction}
The fabrication of materials that are sensitive to physical, chemical, or biological stimuli has opened opportunities for the development of a wide variety of technological applications such as switchable adhesion, mechano-sensing and stimuli-responsive materials \cite{Tawfick_AM2012, LE201813, Chakrapani4009, Boesel_AM2010, Prieto_JAST2016, Prieto_NY2016}. In particular, the design of biomimetic structures \cite{Chakrapani4009, Kwak_SM2010}, inspired by natural systems, has been a powerful tool in the implementation of smart, artificial systems\cite{Trichet_PNAS2012, Liu_NT2011}. In this respect, the use of topographic surfaces is particularly interesting, with natural systems utilizing physical structures, from the nano- to the macro-scale, to deliver functions such as superhydrophobicity, adhesion and anti-biofouling as demonstrated by the lotus leaf, shark skin and gecko feet \cite{Boesel_AM2010, Greener, Kwak_SM2010, Liu_NT2011, Dean4775, Guo_Plant01, Feng001}.

There has been particular interest in developing mechanically responsive systems\cite{Trichet_PNAS2012, Fratzl_N2009}. An excellent example is the mechanical response of micropillar arrays upon drying of water (or water-based solutions) \cite{Bico_N2004, Yeh_SF2012, Nill_ME2011, Pokroy_Sc2009, Chandra_ACR2010, Chandra_L2009, Roman_JPCM2010, Marchand_AMJP2011, Weijs_SM2013, Yang_NP2011, Wei_IPRS2015, C7SM01426K}. When water droplets evaporate on relatively soft elastic microstructured surfaces, capillary action can generate a significant force that is able to bend the soft micropillars. Depending on the geometry of the arrays, the capillary and elastic forces can form different pillar assemblies\cite{Bico_N2004, Yeh_SF2012}. The complexity of the assemblies varies with the pillar height and the inter-pillar distance. For example, large periodic chiral aggregates can be formed when the micropillars are higher and closer to each other. Each cluster of aggregates has a different potential to store elastic energy, embody information, enhance adhesion or capture particles \cite{Nill_ME2011, Pokroy_Sc2009}.

The demonstration of mechanically responsive topographic surfaces to bacterial stimuli during evaporation of small droplets is of great interest and has not been demonstrated before. Furthermore, the deflections seen in our systems are significant leading to pillar aggregations into dimers, trimers, and higher order clusters. Recently, the formation of biofilm strings and networks between topographic pillars has been demonstrated in liquid media,\cite{Zeinab} however, the mechanical response of the pillars to bacterial presence upon evaporation is not observed. Chew and co-authors have shown small deflections of macro-pillared surfaces in response to the differential pressure exerted by biofilm growth within a growth chamber over a 24h period,\cite{Chew2016} while Biais\cite{Biais} and Ng\cite{Dixon} et al. have investigated the interaction of bacterial pili with pillared structures.

Here, we demonstrate how epoxy-made soft surfaces containing micropillar arrays interact with suspensions of different bacterial species. Our results suggest that the presence of motile bacteria with flagella drastically increases the mechanical response of the pillars, actively bending soft topographical substrates in the area contained within the contact line. In contrast, solutions containing non-motile bacteria do not generate such responses. We attribute this to the ability of motile bacteria to interact with each other and with their topographical environment. Importantly, the response of the microarray is sensitive to the type and concentration of bacteria in the solution.These  promising results could lay the foundation for the development of devices that are selectively responsive to specific microorganisms, paving the way to construct smart,  fast  and cost-effective diagnostic tools.

\begin{figure}
\includegraphics[width=0.45\textwidth]{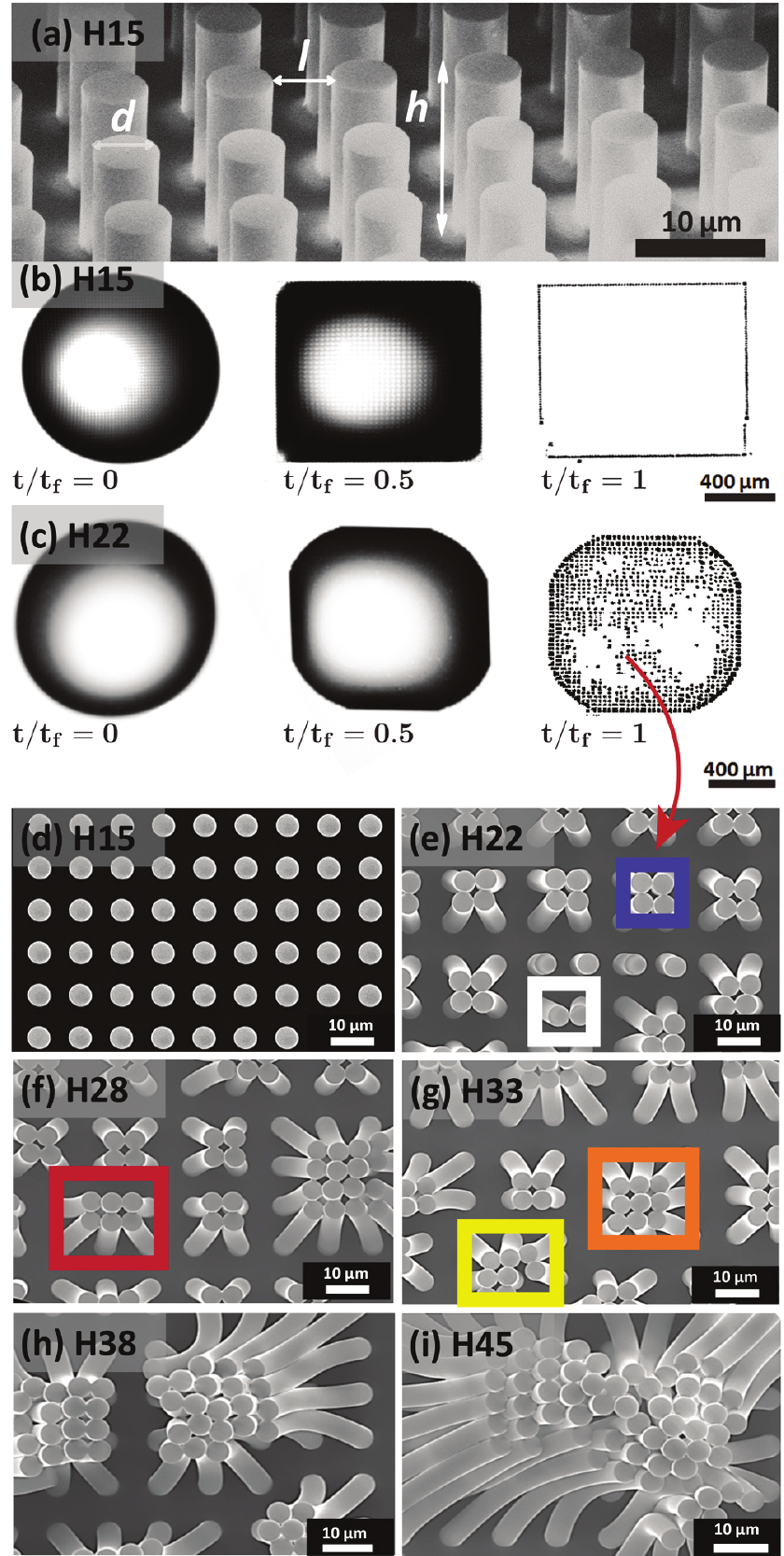}
\caption{(a) Representative SEM image of pillared structure (H15), showing the topographic descriptors for the array. The pillars have a cylindrical shape and a height (h) h = 15 $\mu$m and a diameter (d) of 5 $\mu$m forming a square lattice with an inter-pillar distance l = 5 $\mu$m. (b) Pure water droplet evaporating on the H15 substrate with micropillars leaving a distinct squared shaped contact line with no perturbation of pillars within this contour. (c) Pure water droplet evaporating on the H22 substrate with micropillars leaving a distinct shaped contact line pattern with significant modification of the micropillars within the contact line boundary. Time needed is represented in a dimension less form as the ratio between the elapsed time ($t$) and the final evaporation time ($t_f $). (d-i) Pillared structures with constant (d = 5 $\mu$m), and different pillar heights h of (d) 15 $\mu$m (H15), (e) 22 $\mu$m (H22), (f) 28 $\mu$m (H28), (g) 33 $\mu$m (H33), (h) 38 $\mu$m (H38) , (i) 45 $\mu$m (H45). SEM images are presented for the different heights after evaporation of pure water droplets, probing the sensitivity of the structures to pure elastocapillary bending.}

\label{Fig:1}
\end{figure}

\section{Experiments and Methods}

The epoxy micropillars were fabricated by casting EPO-TEK OG142-13 from Epoxy Technology into a negative replica PDMS mould, as described in \cite{Susarrey_L2016, Hochbaum}. After casting the resin, 1.1 mm thick glass slide is placed over the mould and placed below an ultraviolet light for $20$ min until the epoxy pillar were cured. The epoxy micropillars were mechanically removed from the mould. The SEM images of the epoxy pillars are shown in Figure S1. After the sample preparation, we measure the Young modulus (E) of the bulk material and the micropillar via axial compression test. E for the bulk material was $1\pm0.3$ GPa, and E for the H15 substrate was $0.5\pm0.2$ GPa.

Bacterial cultures were performed following recommended growing conditions for each species. \textit{P. aeruginosa} ATCC-8626, \textit{E. coli} ATCC-10798 and \textit{S. epidermidis} ATTC-12228 were grown over night at 37$^o$C in liquid broth medium (Oxoid, Ltd-Thermo Fisher). \textit{B. subtilis subsp. subtilis} ATCC-6051 and  \textit{L. sakei} DSMZ-20017 were grown overnight at 30$^o$C in MRS broth media from Oxoid Ltd-Thermo Fisher. All the cells cultures were then centrifuged and re-dispersed in sterile deionized water two times, finally adjusting the bacterial concentration to 10$^7$ colony-forming units per milliliter (CFU/mL), unless differently specified. Note that colony counting was performed after cell re-dispersion in deionized water to ensure cell viability.

The evaporation of all droplets was carried out placing a droplet of $5-10\;\mathrm{\mu}$L $\pm\;4\;\mathrm{\mu}$L on the epoxy substrates. For droplets containing bacteria, experiments were performed in triplicates drying 5 droplets over substrates independently. The images were collected with a CMOS camera PCO Sensicam at 1 frames per second (fps). Droplet completely evaporated in approximately $2100\;\pm\;300$ s. Evaporation experiments were assessed at room temperature ($21^o \pm 3^o$ C) in an atmosphere with a relative humidity of $35\;\pm\;5\%$. 

The contact angle measurements of water and bacterial suspension droplets on epoxy surfaces were carried out by placing a water droplet with bacterial suspension of 10$^7$ CFU/mL on the epoxy substrates. The contact angle (CA) for H15 was $100^o\;\pm\;7^o$, whereas the CA was $92^o\;\pm\;5^o$ for H22, H28 and H33. For longer pillars like H38 and H45 the CA was $88^o\;\pm\;3^o$. CA hysteresis were carried out in a similar manner than CA measurements but by tilting the substrate $45^o$. Experiments were performed for H15 substrate with and without bacterial containing droplets only, the CA hysteresis was $50^o \pm 8^o$. No significant differences in CA, and CA hysteresis were observed between water droplets and the deposited bacterial containing droplets. CA values are shown in table S1.

\begin{table*}
  \caption{General characteristics of the different bacterial strains used in the study. AFM images of cells are presented in Figure S2}
  \label{table}
  \begin{tabular}{|l|c|c|c|l|}
    \hline
    \textbf{Strain}  			& \textbf{Gram}		& \textbf{Shape}	& \textbf{$L \times W_a (\mu m^2)$} 	&\textbf{Flagella}  \\  \hline
    (a) \textit{P. aeruginosa}  	& -				& rod			& 1.4($\pm0.2$)$\times$0.8($\pm0.2$) 	& Yes \\
    (b )\textit{E. coli} 			& -  				& rod			& 1.7($\pm0.2$)$\times$0.9($\pm0.2$) 	& Yes\\
    (c) \textit{B. subtilis}		& +  				& rod			& 1.8($\pm0.4$)$\times$0.80($\pm0.2$) 	& Yes\\
    (d)	 \textit{L. sakei}  		& +  				& rod			& 1.5($\pm0.4$)$\times$0.8($\pm0.2$) 	& No\\
    (e) \textit{S. epidermidis}	& + 				& spherical		& 1.3($\pm0.3$)$\times$1.3($\pm0.3$) 	&No\\
    \hline
  \end{tabular}
\end{table*}

Transmission light microscopy images of the dried patterns were collected with a Zeiss 510 confocal microscope equipped with x10, x20 and x40 air objectives. AFM measurements from supporting information were obtained using a Bruker Multimode 8 and a Keysights 5500 instruments. Prior to AFM morphological analysis, a droplet of bacteria suspension (10$^7$ CFU/mL) was deposited onto oxygen plasma treated epoxy flat substrate and dried at room temperature. Estimated length (L) x width (Wa) in Table I are reported within a standard deviation of 10 to 25\% obtained by measured 15 to 20 cells per bacterial strains. These tests were carried out independently in triplicates. Top-view scanning electron microscopy (SEM) imaging was performed at 20 kV. Side-view SEM was recorded after fracturing the epoxy/glass with a diamond cutter at accelerating voltages of 3 kV. Prior to SEM inspection in a JSM-6610 JEOL system, all samples were coated with 20 nm of chromium to increase the electrical conductivity. SEM images are presented without fixation which involves several solvent exchange steps\cite{Susarrey} preserving the bacterial footprints after droplet evaporation.
\section{Results and Discussion}

One of the key parameters in the mechanical response of soft micropillar arrays is the aspect ratio of  single pillar. We investigated the effect of pillar aspect ratio by fabricating regular patterns of cylindrical pillars with a constant diameter (5 $\mu$m) and interspacing (5 $\mu$m), and with variable height (from 5 $\mu$m to 45 $\mu$m). The patterns were created on epoxy resin using a method described before\cite{Susarrey_L2016, Hochbaum, Lix, kim, Pokroy12} based on casting uncured epoxy on a negative polydimethilsiloxane (PDMS) mould, followed by curing and mechanically removing of the mould. The micropatterns were transferred efficiently, with a high degree of fidelity, as shown by scanning electron microscopy (SEM) imaging (Figure \ref{Fig:1}; and Figure S1).  

These microstructured substrates can be susceptible to elastocapillary forces in the presence of pure liquids. Therefore, we evaluated the effect of pure water over a surface decorated with micropillars with lengths varying from 5 $\mu$m to 45 $\mu$m (Figure~\ref{Fig:1}) during the evaporation of water droplets (Figure \ref{Fig:1}). In these experiments, the liquid filled up the space between the pillars resulting in an almost square-shaped droplet contour. Once the droplet spreads on the substrate, the liquid contact line is blocked by the pillared structure and remains immobilized (pinned) for the rest of the drying process.\cite{Susarrey_L2016} Figure~\ref{Fig:1}b shows that after complete evaporation, there is almost no trace of the droplet, except at the droplet contour, where lines of pillars were bent by capillary action at the contact line shown in video S1.\cite{Pokroy_Sc2009, Susarrey_L2016, Chandra_ACR2010, Chandra_L2009, Roman_JPCM2010, Marchand_AMJP2011, Weijs_SM2013}

In the systems studied, the pillar lattice was kept constant (i.e. l = d = 5 $\mu$m) but different pillar heights (h), ranging from h = 5 to 45 $\mu$m were fabricated. Thus a range of micropatterned surfaces were generated with different aspect ratios (i.e. h/d  =  3  to h/d = 9). For large aspect ratio structures, we observed significant perturbation of the micropillars in the area within the contact line boundary. Imaging at low magnifications, or even examination by the naked eye, revealed that the inner part of the pattern was opaque, suggesting that the whole array of pillars inside the dried droplet perimeter was modified (Figure \ref{Fig:1}c). Higher magnification SEM imaging showed that this optical contrast effect was caused by local bending of the micropillars (Figure \ref{Fig:1}d-i), with the pillars bent towards each other forming clusters and adopting complex geometries, e.g. dimer (white box), tetramer (blue box), hexamer (red box), octamer (yellow box), and nonamer (orange box). Similar effects have been reported before for larger pillar aspect ratios \cite{Pokroy_Sc2009, Yang_NP2011, Wei_IPRS2015} and were attributed to the elastocapillary coalescence of the flexible structures \cite{Bico_N2004, Pokroy_Sc2009}. In our experiments, as the aspect ratio decreased, the clusters contained lower numbers of aggregated pillars until a critical aspect ratio h/d = 3 for which no clusters were observed in the inner part of the droplet (Figure \ref{Fig:1}d).

The deformation of the pillars, upon water evaporation, is induced by the surface tension $\gamma$ of the water/air meniscus connecting the pillars, and the corresponding force scales as $F_c \sim \gamma r$, where $r = d/2$ is the pillar radius\cite{Roman_JPCM2010, hadjittofis_2016}. The natural elasticity of the pillars resists deformation with an elastic force $F_E  \sim E l r^4/h^3$, where $E$ is the Young modulus and $l$ the inter-pillar distance.\cite{Pokroy_Sc2009} This expression is analogous to the usual beam theory for slender objects, showing that the resistance to bending decreases strongly when the pillars height increases. If we define the pillar bending sensitivity as the ratio of capillary and elastic forces, $F_c/F_E = \gamma/E l(h/r)^3$,  we can conclude that it is directly proportional to the cubic power of the pillar aspect ratio $h/r$, i.e. slender pillars are more prone to be bent by surface tension, while wide pillars tend to be more stable. 

Under our experimental conditions, no pillar coalescence is observed in the area within the contact line boundary from pure water when the aspect ratio is below $h$/$d$ = 3 \cite{Susarrey_L2016}, suggesting that this is the critical aspect ratio threshold for which capillary action equals restoration mechanical stress on the micropillars. It is important to note that in this analysis, we are not considering the effect of the contact line. This effect is expected to have an enhanced deforming effect but accurate evaluation of this factor is beyond existing phenomenological modelling capabilities and will be the subject of future studies. Consequently, all the results described below applies exclusively to the inner part of the dried pattern left by the droplet, ignoring possible contact-line effects.

\begin{figure}
\includegraphics[width=0.45\textwidth]{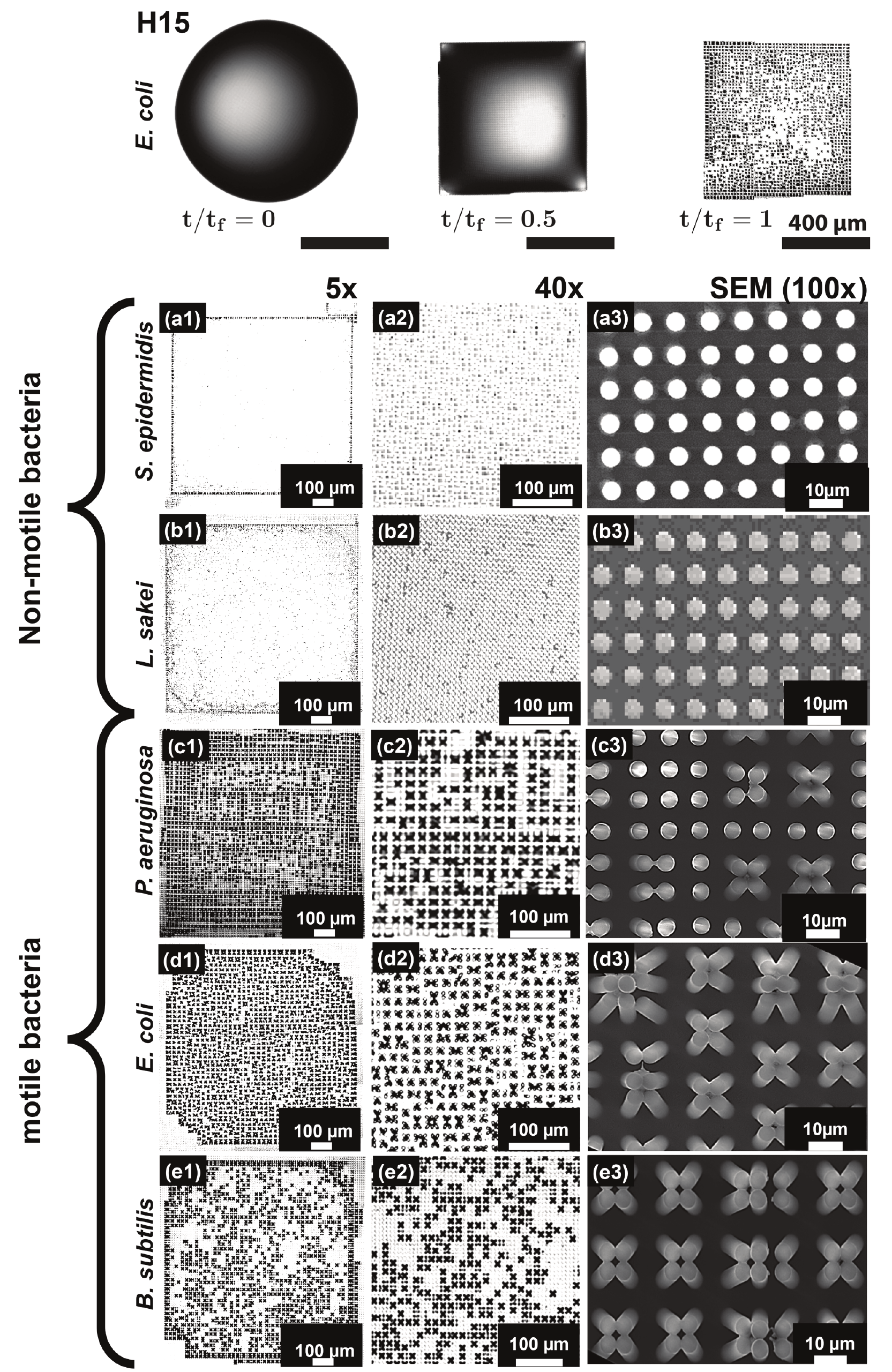}
\caption{Typical patterns left over H15 substrates after the evaporation of different bacterial species: (a1-a3) \textit{S. epidermidis}, (b1-b3) \textit{L. sakei}, (c1-c3) \textit{P. aeruginosa}, (d1-d3) \textit{E. coli}, (e1-e3) \textit{B. subtilis}. Here, the concentration of the different bacterial species is 10$^7$ CFU/ml. The different columns correspond to different degrees of magnifications: 5x (left column), 40x (central column) by using a confocal microscope, and >100x with SEM (right column).}
\label{fig:2}
\end{figure}

\subsection*{Bacterial-triggered coalescence of pillars}

From the elastocapillary assay discussed in the previous section, we identified the critical region within the topographic parameter space where the micropillared structure is able to resist capillary deformation in the presence of pure water droplets. Such a surface opens up the possibility to sense the presence of a second entity introduced into water (i.e. bacterial cells), which could induce a response in its own right. This critical structure corresponds to an aspect ratio $h$/$d \approx$ 3 and pillar height $h$ =15 $\mu$m (H15, Figure~\ref{Fig:1}d), as discussed in the previous section.

We, therefore, investigated the drying process of droplets containing different bacteria species over the H15 pillared structures. Similar to the case of pure water droplets, a pinned square drop shape is found. However, the patterns observed within the contact line formed after complete evaporation of the droplets were surprisingly different for some bacteria as clearly observed in video S2.

Five different bacterial species, with a wide range of morphological and biological characteristics were investigated: \textit{S. epidermidis}, \textit{L. sakei}, \textit{P. aeruginosa}, \textit{E. coli} and \textit{B. subtilis}. The patterns formed after evaporation of droplets containing different bacteria on H15 pillar substrates (Figure~\ref{fig:2}) can be classified in two main groups: one group displaying significant bending of the pillars within the pattern (\textit{P. aeruginosa}, \textit{E. coli} and \textit{B. subtilis}); and, another group which does not induce any responsive bending of the pillars in the center of the dried patterns (\textit{S. epidermidis} and \textit{L. sakei}). These distinct behaviors could be observed even by the naked eye in the form of a local change in contrast at the surface (Figure~\ref{fig:2} 5x). At higher magnifications, the difference is clearly revealed to be associated with the coalescence of adjacent pillars (Figure~\ref{fig:2}, 40x and SEM (100x)).

We attempted to correlate these results to the general characteristics of the bacterial species used in this work (Table~\ref{table}). Atomic force microscopy (AFM) imaging confirmed the expected size and cell morphology for these bacteria: Gram negative (-) \textit{P. aeruginosa} \& \textit{E. coli}, and Gram positive (+) \textit{B. subtilis} \& \textit{L. sakei} present a rod-like shape, while Gram positive (+) \textit{S. epidermidis} have a spheroidal shape (Figure S2). In addition, \textit{L. sakei} and \textit{S. epidermidis} are not motile (no flagella present), while the other three strains have flagella. From these considerations, we can conclude that the different pattern types showed in Figure~\ref{fig:2} (bending vs non-bending) cannot be explained considering bacteria cell morphology only. Similarly, the stiffness of the cell envelop does not appear to play a critical role, with rigid Gram+ bacteria and softer Gram- bacteria distributed among both pattern groups.

\begin{figure*}[t!]
\includegraphics[width=0.75\textwidth]{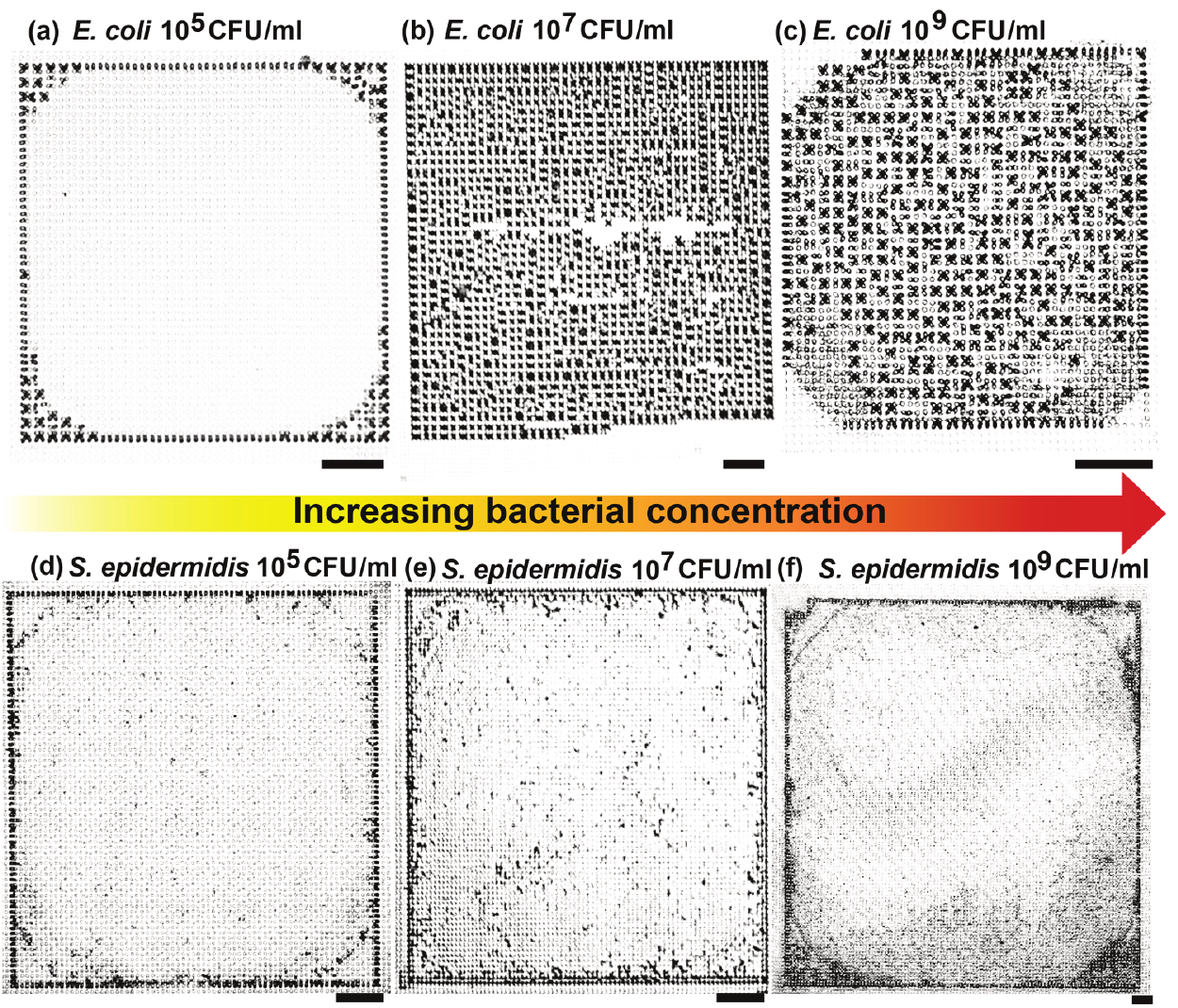}
\caption{Effect of bacteria concentration on the bending pattern for \textit{E. coli} and \textit{S. epidermidis} on H15 pillared substrate. Representative optical microscopy images for a) 10$^5$ CFU/ml; b) 10$^7$ CFU/ml; c) 10$^9$ CFU/ml. Scale bars in (a-f) is 100 $\mu{m}$.}
\label{fig:3}
\end{figure*}

Interestingly, the different response of the microstructures upon evaporation of the bacterial solutions correlates with the presence or absence of flagella. Bacteria with flagella clearly induce a bending response in the H15 pillars, while non-flagellated bacteria are unable to bend the pillars when used at the same bacterial concentration.

For the bacteria that induce a mechanical response, a concentration dependence is observed, with deformation of pillar clusters at the center of the dried droplet observed for bacteria concentrations between 10$^7$ CFU/mL and 10$^9$ CFU/mL, while none is observed for lower bacteria concentrations (10$^5$ CFU/mL). At low concentrations, only the perimeter near the corners of the dried square pattern presented coalescence of the pillars (Figure~\ref{fig:3}a-c). This can be attributed to the coffee-stain-like effect, able to drag bacterial cells towards the droplet contact line, increasing the local concentration of bacteria during evaporation \cite{Susarrey_L2016}. Interestingly, bacterial cells without flagella confirms the absence of responsivity at different cell concentrations (Figure~\ref{fig:3}d-f).

No clear correlation was observed between bacterial species and the cluster symmetries obtained (e.g. dimer, trimer, tetramer, etc.). However, the data suggests that the assemblies emerges due to perturbation of the balance between capillary forces and elastic restoration forces in the presence of bacteria with flagella. In the next section, we discuss a possible mechanism for this distinctive behavior.

\subsection*{Possible origin of bacteria-induced coalescence}

In the previous sections we determined the critical pillar aspect ratio below which surface tension forces were not able to induce pillar coalescence in pure water. Interestingly, the responsivity is dramatically enhanced when the droplets contain flagellated bacteria. While the bending process at the perimeter of the contact line appears similar in both cases, coalescence within the central area is triggered at smaller aspect ratios by the presence of bacteria with flagella. This enhanced pillar-bending effect results in characteristic patterns on the substrate, distinct for motile and non-motile bacteria.

The possible origin of the enhanced pillar-bending may be related to the ability of the bacteria with flagella to adhere to more than one pillar (Figure S3), thus connecting adjacent pillars and inducing a mechanical deformation. In the presence of bacteria with flagella, we observed at SEM, after drying, structures bridging bent pillars, while non-flagellated bacteria appeared attached to single pillars. The morphology of the single bacterial cells cannot be distinguished, probably due to distortions on the cell envelop after evaporation, in the absence of fixation.

These effect can also be understood comparing the length scales of bacterial structures and pillar interspacing distances. The average size of the capsule for a single bacterial cell is below 2 $\mu$m (Table~\ref{table}), while flagella can reach tens of $\mu$m beyond the outer cell membrane.\cite{Haiko} Considering that in our microstructured surfaces the inter pillar distance was 5 $\mu$m, bacteria without flagella will predominantly fall between the pillars or strongly adhere\cite{Hizal} to single pillars. On the other hand, bacteria with flagella,\cite{Hochbaum} in which appendage sizes exceed the inter-pillar distance, can potentially interact with more than one pillar, leading to the observed pillar deformation. 

\begin{figure}
\includegraphics[width=0.45\textwidth]{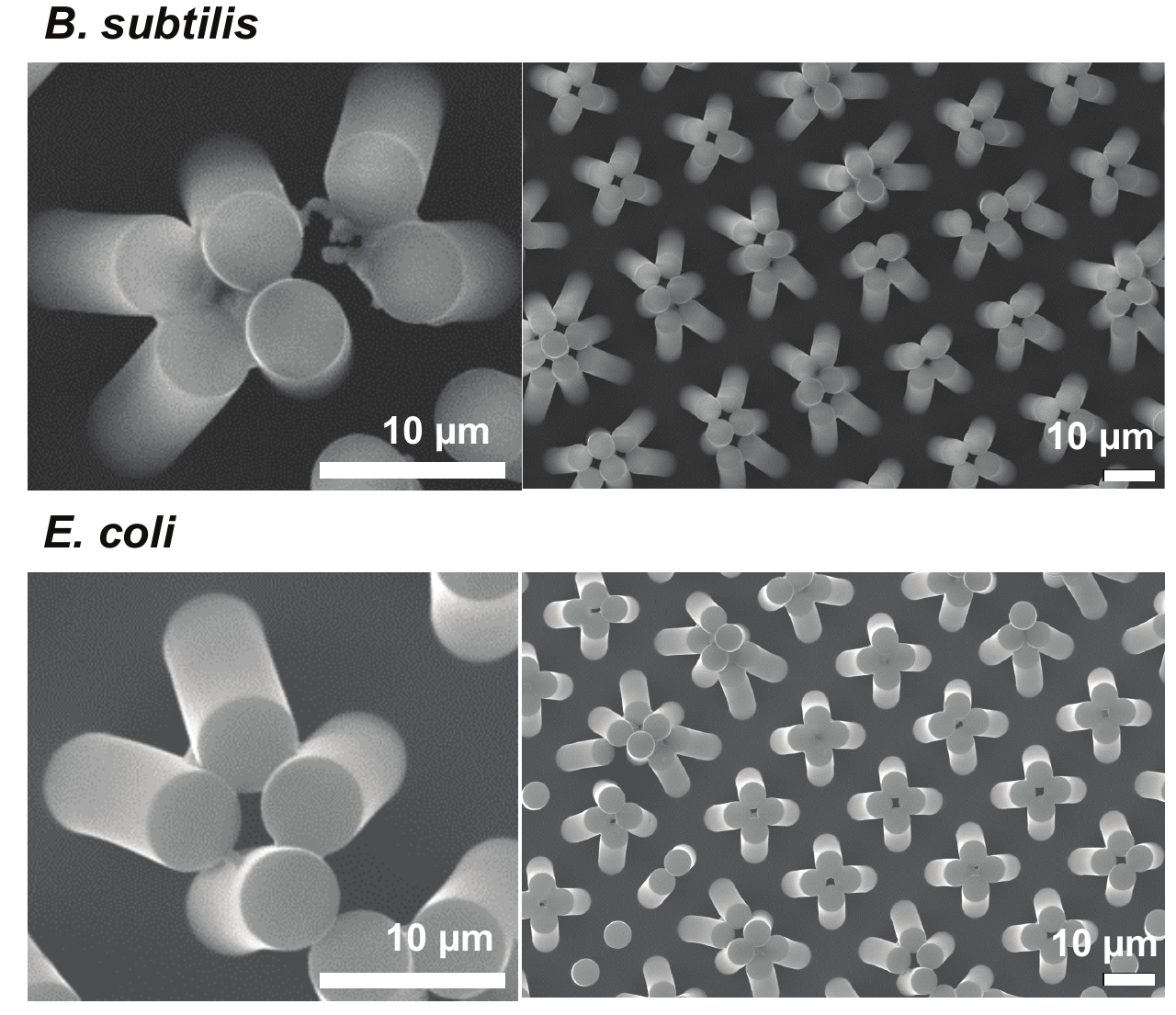}
\caption{Representative SEM images of pillared structures H15 after drying of bacterial suspensions, showing motile bacteria (\textit{B. subtilis} and \textit{E. coli}) bridging the bent pillars. The concentration of the different bacterial species is 10$^7$ CFU/ml.}
\label{fig:4}
\end{figure}

In support of this, we found evidence of bacterial matter residing between the bent pillars, after complete evaporation of droplets containing flagellated bacteria (Figure~\ref{fig:4}). Non- flagellated bacteria, on the other hand, are found attached to individual pillars only, forming non-connecting structures (see Figures S4-S7).

Although a more detailed investigation of bacterial behavior during the actual drying process is necessary to confirm the  hypothesis proposed, our results support the potential use of pillared soft substrates to discriminate between motile and non-flagellated bacteria using a cost-effective and immediate assay based on droplet-drying, that can be performed and quickly analyzed by the naked eye. In addition, discrimination of bacterial concentration is also possible, with only samples containing concentrations above a critical threshold producing a response. We envision that by tuning the properties of the substrates, a more subtle differentiation between different microorganisms and different bacterial concentrations could be achieved in the future with presented here novel, easy to fabricate and cost effective technology.

\section*{Conclusions}
We show that soft micropillared surfaces can be tailor-made sensitive to the presence of isolated bacterial cells in a single drop. The evaporation of water droplets and bacterial suspensions over fabricated micropillar arrays leads to very distinct micropillar deformations and patterns. Once the threshold for elastocapillary pillar coalescence is found, we observe that only bacteria with flagella can promote pillar coalescence. Such responsive micropillared surfaces could provide a platform  for the development of fast and cost-effective self- responsive surfaces for bacterial detection and differentiation.\\

\section*{Acknowledgements}
We would like to thank Dr. Joanna Wnetrzak and the Liverpool Centre for Cell Imaging (CCI) for help with experimental design and technical support. We also acknowledge the support of the Nanoinvestigation Centre at University of Liverpool (NICAL) for access to the SEM facility. Stefan Schlautmann (Mesoscale Chemical Systems, MESA + Institute of Nanotechnology, University of Twente) is also acknowledged for sample fabrication. This work was partly founded by BBSRC (BB/R012415/1).

%

\end{document}